\documentclass[pra,aps,floatfix,tightenlines,showpacs]{revtex4}
\usepackage{amsmath}
\usepackage{amssymb}
\usepackage{amsfonts}
\usepackage{epsfig}
\usepackage{subfigure}
\usepackage[dvips]{color}
\usepackage{bm}

\begin{document}
\title{A Double-Threshold Technique for Fast Time-Correspondence Imaging}
\author{Ming-Fei Li}
\affiliation{
Institute of Physics, Chinese Academy of Sciences, Beijing 100190, China}
\author{Yu-Ran Zhang}
\affiliation{
Institute of Physics, Chinese Academy of Sciences, Beijing 100190, China}
\author{Xue-Feng Liu}
\affiliation{Center for Space Science and Applied Research, Chinese Academy of
Sciences, Beijing 100190, China}
\author{Xu-Ri Yao}
\affiliation{Center for Space Science and Applied Research, Chinese Academy of
Sciences, Beijing 100190, China}
\author{Kai-Hong Luo}
\affiliation{ Institute of Physics, Chinese Academy of Sciences, Beijing
100190, China}
\affiliation{ Integrated Quantum Optics, Applied Physics, University of Paderborn, Warburger Str.~100, D-33098, Paderborn, Germany}
\author{Heng Fan}
\affiliation{ Institute of Physics, Chinese Academy of Sciences, Beijing
100190, China}
\author{Ling-An Wu}
\email{wula@iphy.ac.cn}
\affiliation{
Institute of Physics, Chinese Academy of Sciences, Beijing 100190, China}
\date{\today}
\pacs{42.30.Va, 42.0.Ar, 42.50.St}

\begin{abstract} We present a robust imaging method based on
time-correspondence imaging and normalized ghost imaging (GI) that sets two
thresholds to select the reference frame exposures for image reconstruction.
This double-threshold time-correspondence  imaging protocol always gives better
quality and signal-to-noise ratio than previous GI schemes, and is insensitive
to surrounding noise. Moreover, only simple add and minus operations are
required while less data storage space and computing time are consumed, thus
faster imaging speeds are attainable. The protocol offers a general approach
applicable to all GI techniques, and marks a further step forward towards
real-time practical applications of correlation
imaging.\\$[DOI: 10.1063/1.4832328]$
\end{abstract}
\maketitle

\noindent ``Ghost" imaging (GI) was so named due to its surprising ``nonlocal''
feature, in which the image of an object was retrieved through the second-order
quantum correlation between pairs of photons produced by spontaneous parametric
down-conversion \cite{1994pra, add1995pra}. However, GI with thermal light has also been
demonstrated theoretically and experimentally
\cite{prl2002, prl2004L, prl2005L, prl2005S, add3th2011} and has been interpreted
successfully in terms of classical statistical optics \cite{apl2008, PRA2012}. The
controversy about whether GI is a quantum or classical phenomenon aroused a hot
debate \cite{prl2006, prl2007, addCGIprA2008, addAysSNRprA2008, book2012-1, book2012-2}, but all agree that it
has several advantages for practical applications. (1) Images of an unknown
object may be obtained in a ``nonlocal'' manner, which can be used in remote
sensing \cite{addCGIRS}; (2) the spatial resolution in GI may exceed the diffraction limit \cite{arxiv2009}; (3)
GI can be performed with a true thermal light source without using a lens
\cite{OL2009}, which is useful in x-ray imaging \cite{prl2004H}; (4) imaging is
possible even in a turbulent atmosphere \cite{oe2009, apl2011,addQITurb2,addQITurb} or scattering
medium \cite{OL2011, addDGISM2013} under certain conditions.

However, some limitations still exist; notably, the visibility and
signal-to-noise ratio (SNR) of thermal light GI is low, especially for complex
gray-scale objects. Also, the image cannot be retrieved instantaneously, since
it is based on averaging of the second-order correlation of two measurements,
so a huge amount of data is involved. Happily, these shortcomings are not
unsolvable. The quality can be improved using the techniques of differential
ghost imaging (DGI) \cite{DGI} and normalized ghost imaging (NGI)
\cite{ngi2012}, by means of which the SNR of conventional GI can be enhanced
enormously, although huge amounts of data and more complex computation are
required. Ghost imaging via compressive sensing has been demonstrated which can
give a high SNR with many fewer exposure frames \cite{APLCS1, PRACS2}, but more
computation time is necessary; sparcity constraints also limit its
applications. Thus, less data acquisition, shorter computation times and high
SNR are the ultimate goal of GI. Recently in our group, Luo \emph{et al.}
\cite{cpl2012} reported a so-called correspondence imaging (CI) protocol which
seemed to defy intuition in that no direct second-order correlation calculation
is performed, while compared with conventional GI the number of exposures used
to reconstruct the images and consequently the computation time are greatly
reduced. This was then combined with DGI, i.e. time-correspondence differential
GI, by which means high quality images were retrieved using only part of the
reference detector data without computing the correlation function
\cite{LiPRA}. Although high efficiency and good quality images were obtained by
setting two thresholds, it was not specified how the two thresholds should be
set. In another work, high visibility ghost images were produced using a
synthesized light source with exceptionally large intensity fluctuations,
produced by means of a spatial light modulator \cite{AIPxfl}. Different from
the above two methods, in this paper, instead of modifying the light source, we
employ a special detection system which only picks out the large fluctuation
signals and discards the small fluctuations, thus computational time and
storage space can both be saved.

This approach, which we call double-threshold time-correspondence imaging
(DTTCI), can be applied to all types of correlation imaging, while
incorporating all their advantages. First, the image quality is better than
that of traditional GI. Second, the method is insensitive to surrounding noise.
Third, less exposure frames are needed to recover the image so imaging time and
memory storage are reduced. Fourth, the computer algorithm is simple, so
computation time is saved. With all these advantages, our method should be very
useful in real practical applications.

\begin{figure}[t]
\centerline{\includegraphics[width=8.0cm]{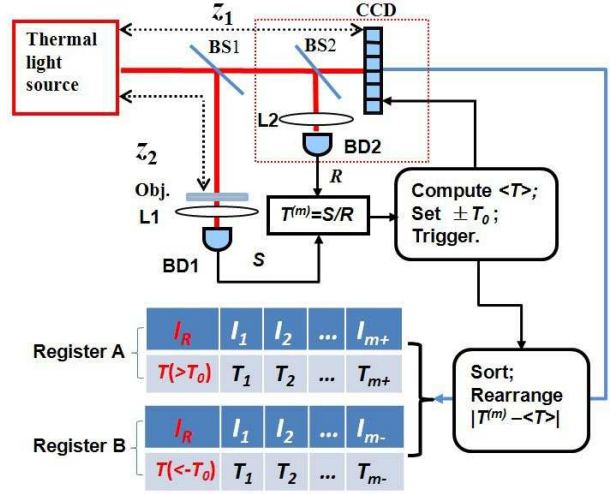}}
\caption{(Color online). Schematic of experimental setup. BS1, BS2
beamsplitters, BD1, BD2 bucket detectors, Obj object, L1, L2 collection lenses,
CCD spatially resolving reference camera, $S$ bucket signal, $R$ total
reference intensity at BD2. A and B are computer registers.\label{setup1}}
\end{figure}

The experimental setup, shown in Fig.\ \ref{setup1}, is a generic lensless GI
system using thermal light, but with a second beamsplitter BS2 and bucket
detector BD2 in the reference beam. The purpose of BD2 is to record the total
intensity arriving at the reference detector CCD, and BS1 is assumed to be a
1:2 beamsplitter, while BS2 is a 50/50 one. The distances $z_1$ and $z_2$ from
the light source to the CCD and object must be equal for lensless GI.

\begin{figure}[h]
\centerline{\includegraphics[width=0.4\textwidth]{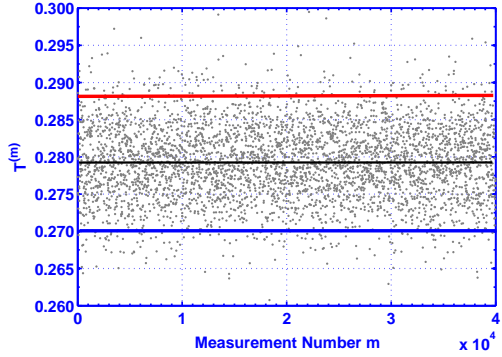}}
\caption{(Color online). $T^{(m)}$ vs. measurement number $m$. \label{setup2}}
\end{figure}

In conventional GI, the image is recovered from the
second-order correlation function of the intensity fluctuations at
two detectors, one of which only measures the total light transmitted through
the object, while the other in an empty reference arm is capable of spatial
resolution:\\
\begin{eqnarray}
GI=\langle\delta S~ \delta
I_{R}(\bm{x}_{R})\rangle\simeq C_{0}T(\bm{x}_{R}),
\end{eqnarray}
where $S$ is the total intensity recorded by the object arm bucket detector, $
I_{R}(\bm{x}_{R})$ is the spatial distribution of the light intensity at the
reference detector, $x_R$ denotes the spatial position, $C_{0}=A_{coh}\langle
I_{R}(\bm{x}_{R})\rangle \langle I_{B}(\bm{x}_{B})\rangle$ is a constant, $
A_{coh}$ is the average speckle size, and $T(\bm{x})$ the intensity
transmission function of the object. The bucket signal $S$ in the object beam
is
defined as \cite{DGI,ngi2012}
\begin{eqnarray}
S=\int_{ } I_{B}(\bm{x}_{B})T(\bm{x}_{B})d^{2}\bm{x}_{B},
\end{eqnarray}
and $\delta S=S-\langle S\rangle$, $\delta
I_{R}(\bm{x}_{R})=I_{R}(\bm{x}_{R})-\langle I_{R}(\bm{x}_{R})\rangle$.
The suffixes $B$ and $R$ denote bucket and reference arms, respectively. We also
define the total reference signal in BD2 to be
\begin{eqnarray}
R=\int_{ } I_{R}(\bm{x}_{R})d^{2}\bm{x}_{R},
\end{eqnarray}
where $\bm{x}_{B}$ and $\bm{x}_{R}$ must be integrated over the same area. With
suitable adjustment of the beamsplitter ratios, $R$ equals the total light
intensity impinging on the object. The total instantaneous transmission
function of the object can be measured as $T^{(m)}=S^{(m)}/R^{(m)}$, where $m$
denotes the $m$-th measurement. Then we obtain the normalized ghost image to be
\begin{align}
NGI &=\langle\frac{S
I_{R}(\bm{x}_{R})}{R}\rangle-\langle\frac{S}{R}\rangle\langle I_{R}(\bm{x}_{R})\rangle\nonumber\\
&=\langle S
I'_{R}(\bm{x}_{R})\rangle-\frac{\langle S\rangle}{\langle R\rangle}\langle R
I'_{R}(\bm{x}_{R})\rangle\nonumber\\
&\simeq C'_{0}\Delta T(\bm{x}_{R}), \label{ngi-dgi}
\end{align}
where $C'_{0}=A_{coh}\langle I'_{R}(\bm{x}_{R})\rangle\langle
I_{B}(\bm{x}_{B})\rangle$ is a constant, $\Delta T(\bm{x})=T(\bm{x})-\langle
T\rangle$, and we have assumed that $\langle T\rangle=\langle
S/R\rangle\simeq\langle S\rangle/\langle R\rangle$. The normalized reference
detector intensity $I'_{R}(\bm{x}_{R})=I_{R}(\bm{x}_{R})/R$ changes as the
contrast of the speckle patterns changes, though its general kurtosis does not
\cite{PRA2012}.
We now tentatively preset two threshold values $\pm T_{0}$ according to the
value of $\langle S/R\rangle$, as illustrated in Fig.\ \ref{setup2}, which is a
plot of $T^{(m)}$ against $m$, when all the frame intensities are recorded. The
two threshold values are represented by the upper (red) and lower (blue) lines.
Only the values of $S/R$ above the upper line or below the lower line are used
to trigger the CCD to take an exposure which will then be saved. Here
$T_{0}\in[0,\max |\delta T^{(m)}|]$, where $\delta T^{(m)}=T^{(m)}-\langle
T^{(m)}\rangle$. Next, according to whether $\delta T^{(m+)}>T_{0}$ or $\delta
T^{(m-)}<-T_{0}$ is satisfied, the corresponding time-correlated frames of the
reference detector will be stored in Register A or B, where $m+$ and $m-$
denote all the frames larger or lower than the selected upper and lower
thresholds, respectively.

Thus we obtain
\begin{eqnarray}
\centering
\langle\frac{\delta T_{0}^{(m\pm)}\delta I_{R}(\bm{x})}{|\delta T_{0}^{(m\pm)}|}\rangle_{k}
=\begin{cases}
\langle\delta I^{(m+)}_{R}(\bm{x})\rangle_{k}, \\ \langle\delta I^{(m-)}_{R}(\bm{x})\rangle_{k},
\end{cases}
\label{pnTk1}
\end{eqnarray}
where
\begin{eqnarray}
&&\delta T_{0}^{(m\pm)}=|\delta T^{(m\pm)}|- T_{0}, \\
&&\langle\delta
I^{(m\pm)}_{R}(\bm{x})\rangle_{k}=\langle
I^{(m\pm)}_{R}(\bm{x})\rangle_{k}-\langle
I_{R}(\bm{x})\rangle,
\end{eqnarray}
and $k$ denotes the number of signals $T^{(m)}-\langle T^{(m)}\rangle$
larger than $T_{0}$ or smaller than $-T_{0}$. Suppose that
\begin{eqnarray}
\langle\frac{\delta T_{0}^{(m\pm)}\delta I_{R}(\bm{x})}{|\delta T_{0}^{(m\pm)}|}\rangle_{k}
\simeq\frac{\langle\delta I_{R}(\bm{x})\delta
T_{0}^{(m\pm)}\rangle_{k}}{\langle|\delta T_{0}^{(m\pm)}|\rangle_{k}}\propto\pm
\Delta T(\bm{x}).
\label{pnTk2}
\end{eqnarray} is satisfied, this requires that $|\delta
T_{0}^{(m\pm)}|\simeq\langle|\delta T_{0}^{(m\pm)}|\rangle_{k}\}$ be satisfied.
It so happens that this condition is indeed satisfied due to the statistical
properties of thermal light. When appropriate values of $T_{0}$ are chosen,
there will be a minimum value of the deviation from the mean of the intensity
fluctuation  $|\delta T_{0}^{(m)}|$, so Eq. (\ref{pnTk2}) is better satisfied
and the corresponding $T_{0}$ provides the optimum data for retrieving the best
quality image, while at the same time both positive and negative differential
images can be obtained.
Thus, according to Eqs. (\ref{ngi-dgi}), (\ref{pnTk1}) and (\ref{pnTk2}), the DTTCI image is given by:
\begin{eqnarray}
\raggedleft
\langle
I^{(m+)}_{R}(\bm{x})\rangle_{k}-\langle
I^{(m-)}_{R}(\bm{x})\rangle_{k}=\frac{{MC'_{0}\Delta
T(\bm{x})}}{
2k{\langle|\delta
T^{(m)}-T_{0}|\rangle}},
\label{p-n}
\end{eqnarray}
where $M$ is the total number of measurements. It is important to note that the
bottleneck in the speed of GI is not only in the post-processing of the images
but in the acquisition of the individual frames in the reference beam. However,
the main time-consuming operation in GI is the latter, as well as the
processing of all the big matrices of the reference CCD frames. The CCD
exposure time is about 100 $\mu$s (10 kHz) but with only about 60 frames per
second sampling rate, while an ordinary photodiode used as the bucket detector
would generally have a response time on the order of ns (GHz). That is why we
add a fast bucket detector in the reference beam to measure the same value of
$R$, instead of using the same CCD to sum all the pixel intensities. This
improvement should be very useful in real applications.

To implement our method, the main difficulty is how to preselect two suitable
intensity threshold values relative to the average of $T^ {(m)}$, since
$\langle T^{(m)} \rangle$ can only be determined after all the values of $T$
have been measured and calculated. Fortunately, we can make a rough estimate of
$\langle T^ {(m)} \rangle$ with only a few hundred measurements, because the
intensity fluctuations of $S/R$ are relatively small and stable for a stable
thermal light source. After the two intensity values $ \pm T_{0}$ have been
preselected, as illustrated in Fig. \ref{setup2}, the CCD camera is triggered
to record only those frames corresponding to fluctuations above $T_{0}$ or
below $-T_{0}$, which will then be stored in registers A and B, respectively.
However, these two sets of data may contain a different number of frames as the
value of $\langle T \rangle$ cannot be known exactly beforehand. This asymmetry
or imbalance will result in poor quality of the image retrieved by using the
subtraction operation in Eq.~(\ref{p-n}). To fix this, our second step is to
rearrange the values of $ T^ {(m)} $ in the register with the greater number of
frames, in ascending or descending order, then certain measurement values
closest to the mean $ T^ {(m)} $ are deleted such that both registers contain
precisely the same number of frames. The third step is to obtain the absolute
difference matrix of these two sets of data.

It should be pointed out that in our method computation time is saved not just
because we only use part of the matrices, but also because we only need to add
the matrices directly, rather than having first to multiply all of the
differential bucket intensity signals one by one with its matching CCD matrix
then integrating, as in the DGI protocol.

In the experiment, a linearly polarized 632.8~nm He-Ne laser beam is projected onto a slowly rotating (0.5~rad/s) ground-glass disk to produce a pseudothermal field of randomly varying speckles. To compare the advantages of DTTCI with other schemes, we perform measurements as in a traditional DGI experiment \cite{DGI}, then employ post-processing algorithms. The thermal beam is divided by a 1:1 beamsplitter (in place of BS1 in Fig.~1, with BS2 removed) into the spatially correlated object and reference beams, which have intensity
distributions $ I_{B}(\bm{x}_{B})$ and $I_{R}(\bm{x}_{R})$, respectively. Both
beams are collected by identical charge-coupled device (CCD) cameras of pixel
size 4.65~$\mu$m. The average area of the speckles at the object and reference
detector planes, which are at the same distance $z_{1}=z_{2}=215$ $\rm{mm}$
from the source. In the experiment, both cameras capture a total of
$8\times10^{4}$ frames, with a synchronized exposure time of 100 $\mu$s. A
digital chart of size $236\times208$ pixels, shown in Fig.~3(a), is
centered on the bucket detector array, from which the total transmitted
intensity $S$ is calculated. The other camera is used as both the bucket
detector BD2 and the reference CCD of Fig. \ref{setup1}, by using
post-processing algorithms.

Experimental results show that the difference $|\langle S/R\rangle-\langle
S\rangle/\langle R\rangle|/\langle S/R\rangle<10^{-3}$ is always satisfied, and
becomes smaller and smaller as the number of measurements $M$ increases. We
also find that the average of $\langle S/R\rangle$ can be predicted with only
about a hundred measurements, for example, $\langle S/R\rangle = 0.2973$ for
both $M=120$ and $M=40000$, as its fluctuations are small for large $M$. The
images obtained by various methods are shown in Fig.~3. Figures
3(b), 3(c) and 3(d) were retrieved from 40000 measurements with
the GI, NGI and DGI algorithms, respectively.

To implement our DTTCI scheme we perform a proof-of-principle simulation
experiment. we designed the specific double threshold values in order to satisfy the very condition through a trial-and-error process. We take $\langle S/R\rangle=0.2973$ as the average value, and set
thresholds $T_{01}=34401\times10^{-7}$ and $-T_{01}=-34742\times10^{-7}$,
for a total of $20,000\times2$ exposures, with 20,000 frames above the positive
threshold and 20,000 frames below the negative threshold. The same number but
not the same 40,000 frames are used in image recovery by the other methods, for
fair comparison. These 40,000 frames are sorted out from a total of
$8\times10^{4}$ exposures, and the recovered ghost image is shown in Fig. 3(e), which is overall better than all the others, with especially much cleaner white parts and smoother transition in the top and bottom gray scales. The results shown in Figs.~3(f) and 3(g) are obtained by setting $T_{02}=34300\times10^{-7}$,
$-T_{02}=-34680\times10^{-7}$, and $T_{03}=66240\times10^{-7}$,
$-T_{03}=-65594\times10^{-7}$, after which $10000\times2$ and $4000\times2$
frames are chosen from a total of $4\times10^{4}$ frames, respectively. Also
using 8000 frames, the image retrieved by NGI is shown in Fig.~3(h), while
that retrieved by DTTCI with $\pm T_{01}$ is shown in Fig.~3(i); again we see that 3(g) and (i) are still somewhat better than (h), with clearer white parts and gray scales. Different pairs of thresholds produce different results. In our experiment the double threshold values were selected by a trial-and-error process to obtain the best images, which were derived by post-processing of the same batch of data using DGI, NGI or GI. To make sure that the numbers of frames of the reference CCD above the threshold $T_{0}$ and below $-T_{0}$ were the same, many threshold values were tried until exactly 20,000 frames both above and below could be obtained from the data. Actually, the choice can be done in more than one way, by arbitrary selection according to the average value of T with software, or preset from experience with hardware.

\begin{figure}[!hbt]
\centerline{\includegraphics[width=0.36\textwidth]{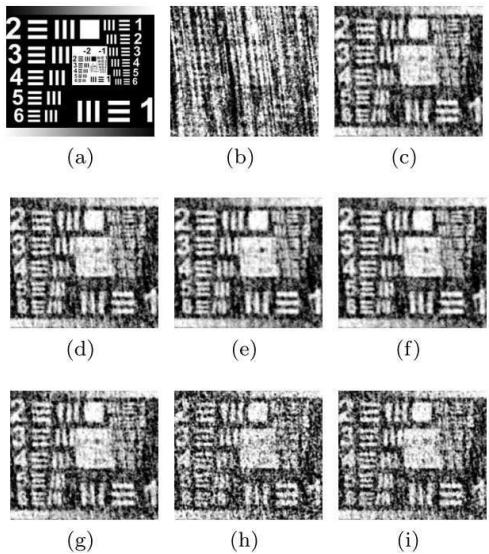}}
\caption{\label{f7} (a) Digital mask Images retrieved from: (b)~GI $40,000$
frames, (c)~NGI $40,000$ frames, (d)~DGI $40,000$ frames, (e)~DTTCI($T_{01}$)
$20,000\times2$ frames, (f)~DTTCI($T_{02}$) $10,000\times2$ frames, (g) DTTCI($T_{03}$) $4,000\times2$ frames, (h)~NGI $8,000$ frames, ~(i)~~DTTCI($T_{01}$) $4,000\times2$ frames.}
 \vspace{-0.0cm}
\end{figure}

To compare the quality of images retrieved with different GI protocols, a
standard normalization method has been used, i.e., the minimum pixel value is
subtracted from each matrix element of the image, then the new matrix is
divided by the maximum pixel value, and the image is normalized to [0,1]. This
is the relative intensity transmission function $t(\bm{x})$ of the object which
we wish to retrieve and which is theoretically equal to the normalized
transmission function. In addition, the Matlab image processing command \cite{histeq}
$``histeq(~)" $ is used to ensure a fair and credible evaluation of the SNR of
all the retrieved images.
\\

\begin{figure}[t]
\centering
{\includegraphics[width=6.5cm]{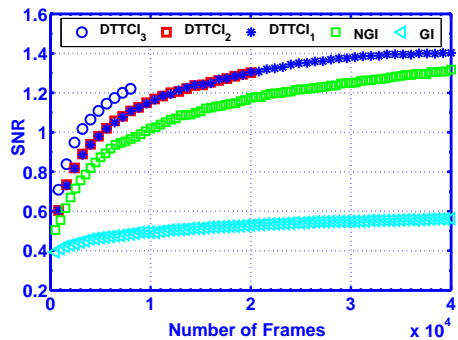}}
\caption{\label{SNRF}(Color online). SNR
vs. number of reference frames.}
\end{figure}

For a quantitative comparison of the image quality, we define the SNR as \cite{LiPRA}:
\begin{eqnarray}
{\rm {SNR}=\frac{Signal}{Noise}}=
\frac{\sum_{i,j=1}^{M,N}[T_{0}(i,j)-\overline{T}_{0}]^2}{\sum_{i,j=1}^{M,N}[T(i,j)-T_{0}(i,j)]^2},
\label{SNR}
\end{eqnarray}
where $T_{0}(i,j)$ and $T(i,j)$ are the transmission matrices of the object
mask of size $M \times N$ and the retrieved image, respectively, and
$\overline{T}_{0}=(MN)^{-1}\sum_{i,j=1}^{M,N}T_{0}(i,j)$. The SNR values for different reconstruction methods vs. the total number of measurements, as shown in Fig.~\ref{SNRF}. The SNRs of DGI and NGI are always better than GI, which agrees with the results of Ref.~27. In all
the DTTCI experiments of Fig.~\ref{f7}, the SNR values are larger and increase
even faster than for NGI. Thus we can say that our method is better than both
DGI and NGI protocols, while the image quality will increase with increase of
the threshold values, and with increase of the number of measurements under the
same threshold conditions.

In conclusion, we have demonstrated both theoretically and experimentally a
double-threshold triggered time-correspondence imaging scheme which is
insensitive to surrounding noise and has a higher SNR than that of previous GI
techniques. The technique is simple and the number of reference frames required
to retrieve an image is greatly reduced, so data acquisition time is shortened.
Moreover, the data processing algorithms are simpler than for DGI or NGI,
consume less computing time, and require less memory storage capacity, without
sacrificing image quality. With hardware implemented for triggering the
reference detector at the thresholds, the total exposure and processing time
could be even shorter. This new protocol offers a general approach applicable
to all GI techniques, and represents a step forward towards real-time
application of correlation imaging.

This work was supported by the National Basic Research Program of China (Grant
No.2010CB922904), the National Natural Science Foundation of China (Grant
No.60978002), and the Hi-Tech Research and Development Program of China (Grant
No.2011AA120102).
\\

\end{document}